# Leveraging Neural Networks with Attention Mechanism for High-Order Accuracy in Charge Density in Particle-in-Cell Simulation

Jian-Nan Chen, Jun-Jie Zhang

*Abstract*—In this research, we introduce an innovative three-network architecture that comprises an encoder-decoder framework with an attention mechanism. The architecture comprises a 1st-order-pre-trainer, a 2nd-order-improver, and a discriminator network, designed to boost the order accuracy of charge density in Particle-In-Cell (PIC) simulations. We acquire our training data from our self-developed 3-D PIC code, JefiPIC. The training procedure starts with the 1st-order-pre-trainer, which is trained on a large dataset to predict charge densities based on the provided particle positions. Subsequently, the 1st-order-pre-trainer, whose predictions then serve as inputs to the 2nd-order-improver. Meanwhile, we train the 2nd-order-improver and discriminator network using a smaller volume of 2nd-order data, thereby achieving to generate charge density with 2nd-order accuracy. In the concluding phase, we replace JefiPIC's conventional particle interpolation process with our trained neural network. Our results demonstrate that the neural network-enhanced PIC simulation can effectively simulate plasmas with 2nd-order accuracy. This highlights the advantage of our proposed neural network: it can achieve higher-accuracy data with fewer real labels.

## I. INTRODUCTION

PARTICLE-IN-CELL method, introduced in the 1960s, is one of the most potent computational tools for studying plasma physics [1]. Using macro-particles to represent ions or electrons within the same phase space, the PIC method self-consistently computes the dynamics of particle motions and electromagnetic (EM) fields [2]-[4], providing an intuitive understanding of the physical attributes and aiding researchers in analyzing plasma phenomena and data.

Several numerical commercial software and open-source codes such as MAGIC [5], ICEPIC [6], UNIPIC [7], Smilei [8], and EPOCH [9], have emerged. These tools are extensively used in designing vacuum electronic devices [10], simulating nuclear explosions [11]-[13], researching space physics effects [14], and studying controlled fusion [15].

However, the efficiency of PIC is a challenge. The ability to handle a larger number of particles leads to the emergency of parallelization. Moreover, the synchronous calculation of electromagnetic fields forms a considerable portion of the total PIC execution time. To address this, recent efforts have aimed at optimizing numerical algorithms, such as merging with fluid-kinetic methods [16] or implicit methods [17].

The order of particle interpolation also impacts the accuracy of PIC. While the grid spacing constrains the order of the Debye length in high-density plasma, high-order spatial interpolation [18] alleviates this limitation and reduces aliasing error [19]. However, an increase in interpolation order significantly escalates computational resource requirements and algorithm complexity, posing challenges in implementation, particularly in 3-D models.

The rise of artificial intelligence (AI) and graphics processing units (GPUs) has led to deep neural networks (NNs) becoming a powerful tool for scientific research, especially in solving partial differential equations in fluid dynamics. In the plasma field, machine learning has been successfully applied to accelerate electrostatic PIC plasma simulations [21]-[23]. NNs operating on GPUs can process hundreds of millions of inputs simultaneously, making them naturally suited for handling the volume of particles in PIC in parallel.

The ATTENTION MECHANISM, one of the most successful neural networks in recent years, captures both local and global connections among inputs. This feature makes it easier to fit and learn long-range connections or dependencies compared to other neural network algorithms. Introduced by the Google machine translation team in 2014 [24] and later advanced in the realm of natural language processing in 2017 [25], the attention mechanism offers advantages like high parallel execution, fewer parameter quantities, and high efficiency and interpretability. The attention mechanism guides the network on what and how much to focus on, allowing it to disregard the distance of input elements.

In this study, we leverage the attention mechanism to learn the relationship between particles and grids, effectively replacing the charged particle interpolation process in the PIC method. Through fine-tuning with fewer high-order labels and a discriminator network, we generate high-order particle interpolation results.

The paper is structured as follows: Section II provides a brief introduction to the fundamental computational process of PIC and the attention mechanism. Section III details how we generate the training data and construct the attention network. Section IV presents the results of charge densities from the neural network and showcases the electromagnetic fields calculated by both the standalone PIC code and the neural-

Manuscript received XX. This work was supported by the National Natural Science Foundation of China under No.12105227 and the National Key Research and Development Program of China under Grant No. 2020YFA0709800. (*corresponding author: Jun-Jie Zhang*).

Jian-Nan Chen and Jun-Jie Zhang are with the Northwest Institute of Nuclear Technology, Xi'an 710024, China and also with National Key Laboratory of Intense Pulsed Radiation Simulation and Effect, Xi'an 710024, China (e-mail: chenjiannan@nint.ac.cn; zhangjunjie@nint.ac.cn).

network-enhanced PIC. Section V concludes the study and provides insights into the potential application of the attention mechanism in the fields of plasma and nuclear physics.

Our work holds the potential to enhance the accuracy of existing codes, and offers a pathway to cross-validation among peers in the field. It suggests that by providing only a portion of the data, one could potentially reproduce others' work on their own programs.

## II. BACKGROUND

### A. Introduction to PIC method

The traditional PIC method comprises four key steps: 1) advancing charged particles, 2) interpolating charged particles, 3) computing electromagnetic (EM) fields, and 4) interpolating EM fields, as illustrated in Figure 1. The first and third step involve solving Newton-Lorentz equations and Maxwell's equations respectively. These equations enable the self-consistent calculation of the EM field and particle distribution by cycling through the steps outlined in Fig. 1.

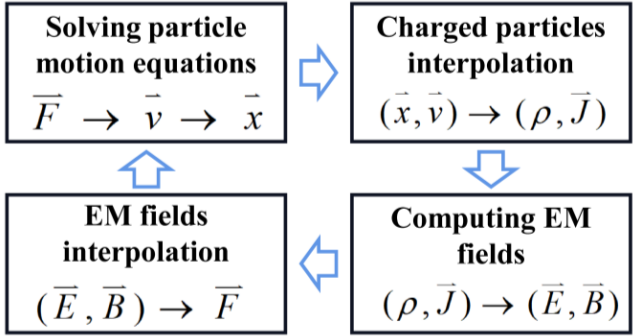

Fig. 1. The traditional particle-in-cell computation process. Order: from top-left to bottom-left.

### B. Introduction to attention mechanism

The attention mechanism is originally introduced via the transformer architecture, as shown in Fig. 2 [25] . This architecture builds upon the encoder-decoder model and incorporates a suite of techniques including residual networks, layer normalization, positional encoding, and masked multi-head attention.

The encoder is composed of a stack of N identical layers, with each layer containing two sub-layers. The first sub-layer performs the multi-head self-attention mechanism, and the second layer contains a simple fully connected feed-forward network. The residual network (Resnet) followed by layer normalization is adopted after the two sub-layers, which prevents the NN from gradient vanishing.

Similar to the encoder, the decoder is comprised of N identical layers. However, it includes an additional third sub-layer sandwiched between the two sub-layers in the encoder. This third sub-layer manages attention over the output of the encoder's hidden information. The decoder also utilizes residual networks and layer normalization, just like the encoder. This symmetrical structure, while seemingly simple, enables powerful and complex computations, central to the effectiveness of the transformer architecture.

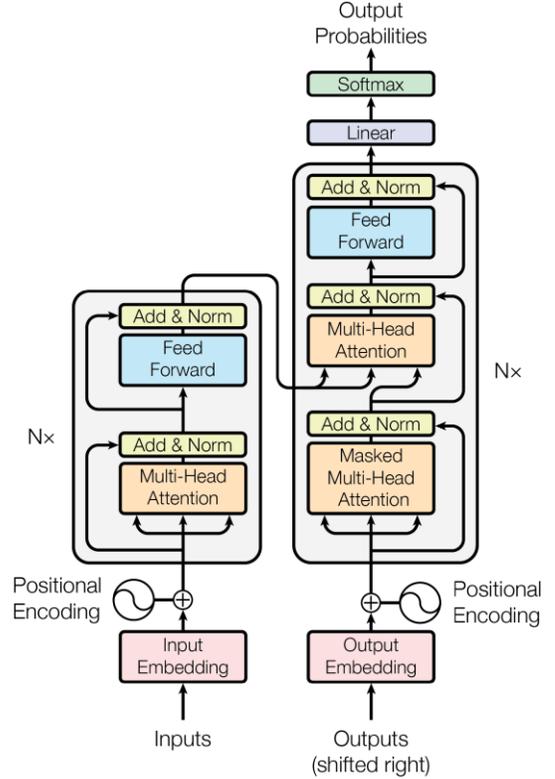

Fig. 2. The transformer model architecture in Ref. [25] .

At the heart of the transformer architecture is the attention mechanism. Its role is to map a query with respective key-value pairs to an output. This output is calculated as a weighted sum of the values, with each value's weight determined by the similarity between the query and its corresponding key. These queries, keys, values, and outputs are all vectors.

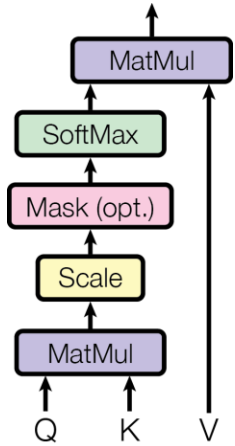

Fig. 3. Scaled dot-product attention in Ref. [25] .

As shown in Fig. 3, the scaled dot-attention model is commonly used. The last dimensions of query and key are both $d_k$, while the value has a last dimension of $d_v$. To derive the weights for the values, we first calculate the dot products (similarity scores) of the query with the keys, then divide each of these scores by $\sqrt{d_k}$, and then apply to a softmax function.

The activation softmax function helps the neural network learn the nonlinear relationship, and the scaling factor prevents the gradients from vanishing and helps the neural network converge faster.

In the execution of the neural network, all the queries are grouped together into a matrix $Q$. Similarly, keys and values are amassed into matrices $K$ and $V$. Then, we can compute the output through matrix operation in Eq. (1):

$$\text{Attention}(Q, K, V) = \text{softmax}(\frac{QK}{\sqrt{d_k}})V \qquad (1)$$

## III. STRUCTURE OF THE PROPOSED NEURAL NETWORK

### A. The preparation of the training data

In this study, we use the 3D fully electromagnetic particle code, JefiPIC [26] , to generate our training and test datasets. We set the total computational region to $20 \cdot dx \times 20 \cdot dy \times 20 \cdot dz$, where $dx = dy = dz = \delta = 10^{-5}$ m. Electron is the only kind of charged particle, with a total charge of $-5 \times 10^{-14}$ C. We use macro-particles to represent the electrons in similar phase space to reduce the computational requirements, leading to the total 20480 particles. The initial particle condition is set that all particles are positioned in a cuboid box with a length of $2 \cdot \delta$, a width of $2 \cdot \delta$ and a height of $10 \cdot \delta$, and they have no initial kinetic energy. Then we tracked the evolution of the charged particle distribution under their self-fields in a time duration of 0.5 ns. To simplify the training process for the neural network, we set a particle reverse boundary to keep the number of particles constant at 20480 throughout the simulation.

All the training and test data are derived from the same initial condition mentioned above. We compute them using $1^{st}$ and $2^{nd}$ order interpolations and sample them at different moments. The "order" refers to the field and charge interpolation order used in JefiPIC. Each dataset includes the grid coordinates, particle coordinates, and particle charges as input, and the charge density on the grid as the label for the network. So we have constructed the training and test data with pairs (input, label) = ([grid coordinates, particle coordinates, and particle charges], charge density on the grid).

We have generated four sets of training data, each referred to as 100-$1^{st}$, 100-$2^{nd}$, 1000-$1^{st}$, and 1000-$2^{nd}$. These names indicate the number of sampled moments and the order of interpolation. The sampled moments of the smaller training sets (100-$1^{st}$ and 100-$2^{nd}$) are included within the 1000 moments of the larger training sets.

The test data consists of two sets, each referred to as the 10-$1^{st}$ and 10-$2^{nd}$ data pairs, with ten moments each.

The actual numerical values of the physical quantities cover a large range and are not suitable for training the neural network. Therefore, we need to normalize all the physical quantities to produce non-dimensional physical quantities, as shown in Table I. The particle charge and grid size are normalized to the unit, while the charge density is normalized by its maximum absolute value. This normalization process helps to scale down the wide range of values to a manageable size, aiding the network to converge more quickly and leading to more effective and efficient training.

*Remark: if the spatial charge density exhibits violent fluctuations, power-law exponentiation can be employed to smooth out the distribution. This technique can help in reducing extreme variations in data, making it more manageable and easier for the neural network to process and learn.*

TABLE I
PEAK VALUE OF FIELD BY DIFFERENT INTERPOLATION ORDERS.

| Physical quantity | Numerical value | Normalized value |
|---|---|---|
| Particle charge | $q_0 = -2.44 \times 10^{-18}$ | $q_n = 1$ |
| Grid size | $\delta = 1 \times 10^{-5}$ | $\delta_n = 1$ |
| Charge density | $\rho$ | $\rho_n = \rho / |\rho|_{max}$ |

### B. The Network Structure

The attention mechanism is known for its ability to learn long-range dependencies among inputs. Therefore, we employ it to capture the allocation coefficient between particles and grids during the particle interpolation process in the PIC method.

The entire neural network proposed in this study, as illustrated in Fig. 4, includes two training processes: pre-training (a) and all-process training (b).

(a) During the pre-training process, we pre-train a basic transformer network, referred to as the $1^{st}$-order-pre-trainer (displayed in Fig. 5), using the larger 1000-$1^{st}$ dataset.

(b) In the all-process training phase, three networks are trained. The first is the same $1^{st}$-order-pre-trainer used in the pre-training stage. The second is a new transformer network, referred to as the $2^{nd}$-order-improver, designed to generate a second-order charge density based on the output from the first transformer. The $2^{nd}$-order-improver is trained using a small number of $2^{nd}$ order charge densities as labels (from the 100-$2^{nd}$ dataset). The third network is a discriminator network (D), designed to help the two transformers perform better.

We will now introduce the structures of these three neural networks in detail.

#### B1. 1st-order-pre-trainer

The first network is $1^{st}$-order-pre-trainer, whose architecture adopted is shown in Fig. 5. In contrast to natural language processing projects, our model removes the masked block and positional encoding modules. Initially, the grid coordinate, particle coordinate, and particle charge are processed by three independent self-attention layers. The outputs from these layers are labeled as the hidden information of the grid coordinate, the particle coordinate, and the particle charge, i.e., query, key and value separately. These query, key, and value are fed into the decoder. After the output passes through a linear layer, the final charge density is generated.

It should be noted that leaky ReLU activation functions always follow the fully connected feed-forward networks. The layer normalization transforms small values into negative numbers with large absolute values, while the leaky ReLU activation function converts all the negative numbers to values close to zero. The combined operation of normalization and leaky ReLU is essentially equivalent to minimizing the impact of smaller connections and enhancing the influence of larger ones from attention.

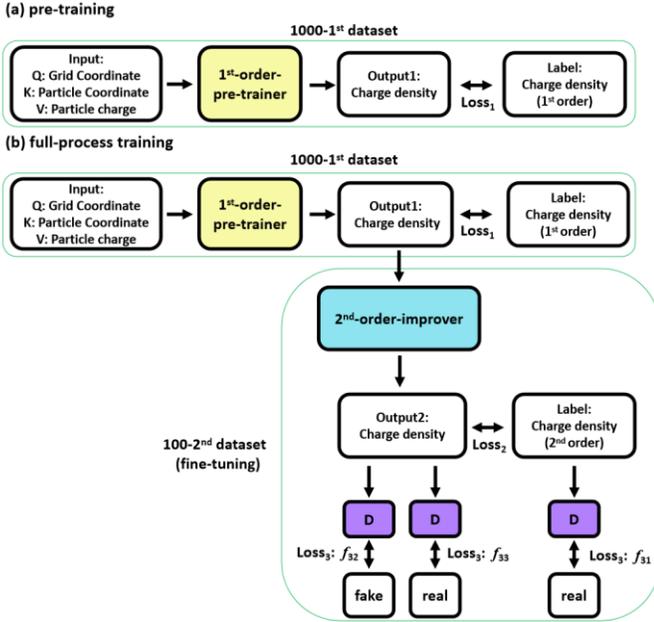

Fig. 4. The designed pre-training and fine-tuning processes. There are three neural networks to train, namely a 1st-order-pre-trainer, a 2nd-order-improver and a discriminator.

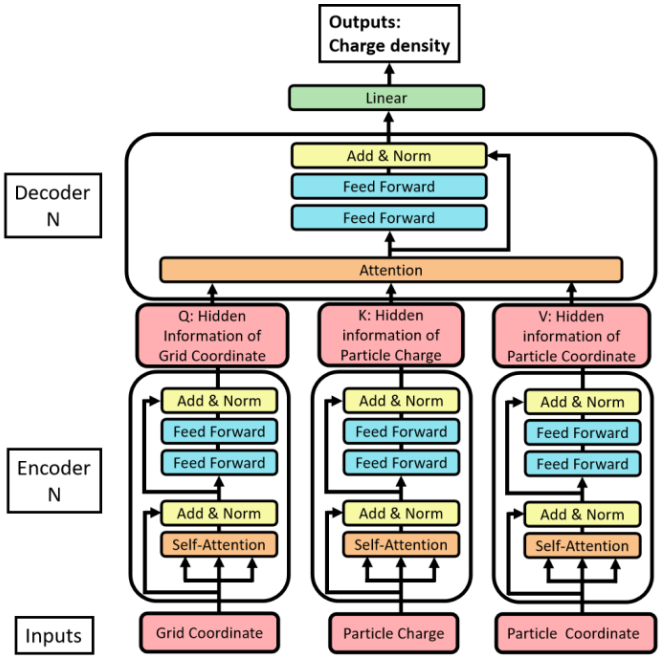

Fig. 5. The model architecture of 1st-order-pre-trainer in our work.

The attention mechanism in the transformer includes both self-attention and attention. The self-attention is used in the encoder to reconstruct the inherent connections of the input itself, where the query, key, and value are all the same input denoted as 'X' in Fig. 6 (a). The attention receives three different inputs, namely the hidden information of the grid coordinate, particle coordinate, and particle charge. It learns the similarity score between the query and key, and generates the charge density, shown in Fig. 6 (b).

In Fig. 6 (b), the front linear layer of the attention mechanism transforms the inputs from a low-dimensional space into a high-dimensional one. The hidden input Q has a shape of (20,20,20,3), where these values represent x-grid size, y-grid size, z-grid size, and spatial dimension, respectively. This input is transformed into a shape of (20,20,20,16) using a fully connected linear layer of shape (3,16). Similarly, the particle coordinate and particle charge are also transformed. The particle coordinate changes from a shape of (20480,3) to (20480,16), and the particle charge from (20480,1) to (20480,16). Here, 20480 stands for the number of particles.

The matrix multiplication of the query and key results in a shape of (20,20,20,20480). The softmax function is performed along the first three spatial axes, summing the output to 20480, which naturally conserves the charge.

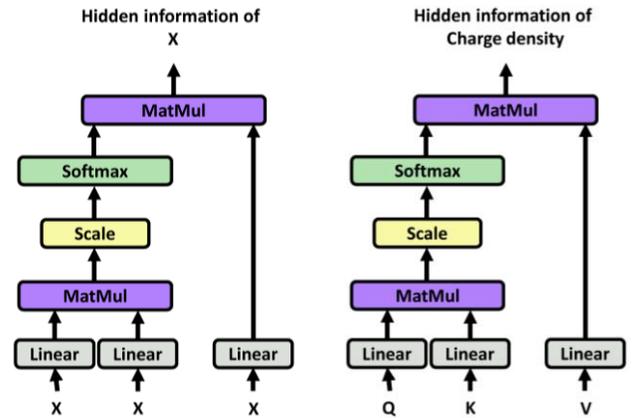

Fig. 6. Structure of attention mechanism. (a) The self-attention process, which is designed to search the long-range dependencies between each element in the same input. (b) The attention process, which is designed to learn the relation or similarity between the query and key.

The loss function of this network is the deviation between output charge density and the 1st order label (from 1000-1st dataset), where $f_1$ is the Huber function. The Huber function is used in this context due to its robustness to outliers, which can help to improve the performance of the model. We use adaptive moment estimation (Adam) optimizer to minimize the loss and update the weights of 1st order transformer through backpropagation.

$$Loss_1 = f_1(Output1,\ 1^{st}\ order\ label) \qquad (1)$$

### B2. 2nd-order-improver

The second network, referred to as the 2nd-order-improver, is designed to improve the order of the charge densities. The design of this network is straightforward, functioning as a self-attention transformer where the query, key, and value are all provided by the output1 from the 1st-order-pre-trainer. This network takes the predicted 1st order charge density, reconstructs it, and forms a new one that is guided by the 2nd order label (in 100-2nd dataset).

Following the pre-training process depicted in Fig. 4(a), the 2nd-order-improver takes output1 as input and generates a new output, output2, which is expected to be of 2nd order. The loss function for this network is the deviation between output2 and the 2nd order label, with $f_2$ being the Huber function.

This deviation is then used in the backpropagation process to simultaneously update the weights of both the 2nd-order-improver and the 1st-order-pre-trainer. This complex process ensures that the model is continuously improving and refining its predictions.

$$Loss_2 = f_2(Output2, \ 2^{nd} \ order \ label) \tag{2}$$

*Remark: 2nd-order-improver is just a fine-tuning operation, and the size of the training dataset (100-2nd) is 10 times smaller than that of 1000-1st. As a result, during this training process, we require that 1000-1st covers 100-2nd.*

### B3. Discriminator network

The first two networks function as the generator, and the third network serves as a discriminator (D). When combined, these three networks form a patchGAN [27] [26] , a type of model that has shown effectiveness in the field of image generation.

The role of the discriminator is to assist the generator in producing more accurate outputs. As shown in Fig. 7, the discriminator is built on a 3-D convolution neural network (CNN).

The discriminator is composed of four CNN layers, each followed by a leaky relu activation function. These components allow the discriminator to capture the features of the input and increase its dimensionality. Following these layers, a fully connected linear layer is used to reduce the dimensionality. This structure enables the discriminator to effectively evaluate the output of the generator and guide the latter to learn.

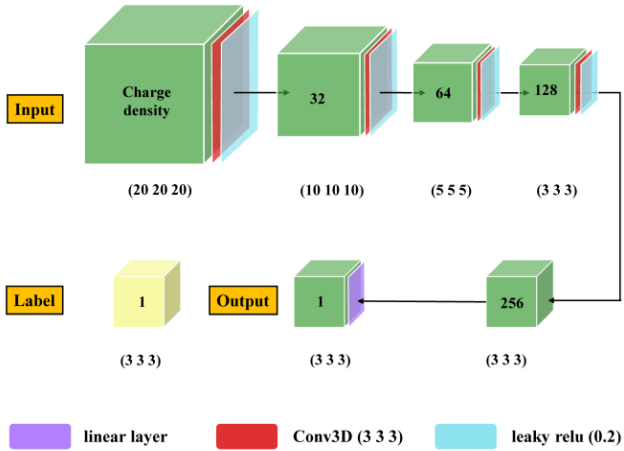

Fig. 7. Structure of patch GAN based on convolution network. The size of the input charge density is (20,20,20,1). The convolution kernel of each layer is (3,3,3). The size of the label is a grid of shape (3,3,3,1).

The discriminator's role is to classify the 2nd order label as 'real' ($Loss_{31}$) and the generator's output (output2) as 'fake' ($Loss_{32}$). To enhance the accuracy of the generator, we also aim to have the discriminator classify output2 as 'real' ($Loss_{33}$) simultaneously. Therefore, the loss function is composed of three parts. The functions $f_{31}$ $f_{32}$ and $f_{33}$ are all Huber functions.

$$\begin{aligned} Loss_3 = &f_{31}(D(2^{nd} \ order \ label), \ real) + \\ &f_{32}(D(Output2), \ fake) + \\ &f_{33}(D(Output2), \ real) \end{aligned} \tag{3}$$

## IV. RESULT

The training of the networks is performed on a single A100 GPU card. To demonstrate the performance of our designed neural networks, which are used to predict the charge density, we select a specific moment from the test datasets displayed in Fig. 8.

Fig. 8 presents the following:

Fig. 8 (a) displays the predicted charge density generated only by the 1st-order-pre-trainer trained after 20,000 epochs.

Fig. 8 (b) shows the 1st order test label, which was computed by the 1st order particle and field interpolation method via JefiPIC.

Fig. 8 (c) presents the predicted charge density generated by the full-process after 25,000 epochs.

Fig. 8 (d) shows the 2nd order test label, which was computed using the 2nd order particle and field interpolation method via JefiPIC.

These figures illustrate the capability of the neural networks to accurately predict charge densities. The close resemblance between the predicted charge densities and the test labels indicates that the networks have learned the intricacies of the particle and field interpolation methods and can successfully generate accurate predictions.

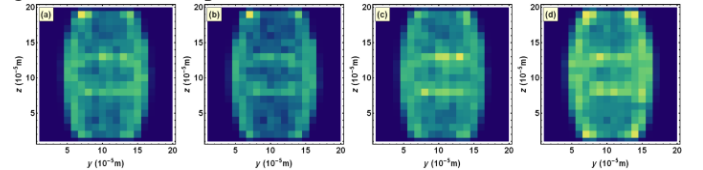

Fig. 8. Charge density distribution. (a) output1 from 1st-order-pre-trainer with 10-1st test dataset. (b) 1st order label in 10-1st dataset. (c) output2 from 2nd-order-improver with 10-2nd test dataset. (d) 2nd order label in 10-2nd dataset.

To evaluate and compare the performance of the two neural networks, the 2nd order label is used as the baseline. The deviations of the predicted charge densities from this baseline, generated by each of the networks, are tabulated and presented in Table II. It is revealed that our proposed full-process 2nd order network performs superior results almost approaching the label.

TABLE II
DEVIATIONS OF THE CHARGE DENSITIES WITH RESPECT TO 2ND ORDER LABEL.

| Charge density | Deviation from 2ND ORDER LABEL |
|---|---|
| output1 (1st-order-pre-trainer) | 9% |
| 1st label | 14% |
| Output2 (2nd-order-improver) | 4% |

It's important to note that we have already accounted for and discounted the statistical fluctuations inherent in the JefiPIC calculations. This means that any deviations observed are largely attributable to the performance of the neural networks, rather than variabilities in the baseline computations. We have made this adjustment to ensure a more accurate and fair comparison of the neural networks' performances.

To further explore and understand the role of the attention mechanism in particle interpolation, we use a simpler case: a single-particle model. In this model, there is only one particle confined in a $10\ \delta \times 10\ \delta \times 10\ \delta$ box with a reverse boundary condition. The evolution data for this model is provided by JefiPIC. We have selected 20,000 moments data as the training data pairs and trained the networks over 200,000 epochs. This simpler case allows us to focus more closely on the attention mechanism by minimizing other variables and complexities. By observing how the network learns the weight allocation in this single-particle scenario, we can gain insights into how the attention mechanism contributes to the process of particle interpolation.

Fig. 9 provides a visualization of the particle density of a single particle on the YOZ plane at $x = 7 \cdot \delta$ for a specific moment. The particle's coordinate is at grid $(7 \cdot \delta, 8 \cdot \delta, 7 \cdot \delta)$. The attention similarity score and the predicted particle density are shown in Fig. 9 (a) and Fig. 9 (b). The particle density from JefiPIC is shown in Fig. 9 (c).

The results predicted by the neural network align well with those computed by JefiPIC, indicating that the model has been successful in its learning process. This suggests that the attention mechanism has effectively learned the weight coefficient of the charge-grid interpolation from the provided data. This consistency validates the effectiveness of the attention mechanism in the context of particle interpolation.

*Remark: Ensuring accurate results necessitates the comprehensive training of all network parameters and the similarity score of the matrix using a considerably large dataset. However, this requirement for extensive data can be viewed as an intrinsic limitation of the attention mechanism, underscoring a trade-off between precision and computational efficiency.*

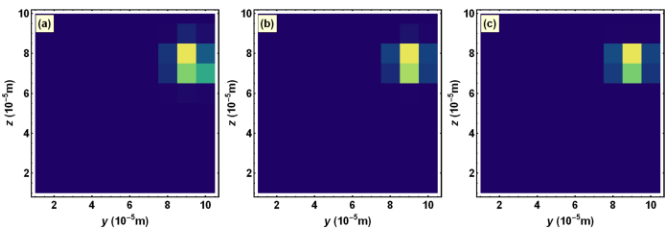

Fig. 9. Results of the single-particle model. (a) Attention similarity score computed by neural network. (b) Particle density computed by neural network. (c) Particle density computed by $2^{nd}$ order interpolation JefiPIC.

### A. Why such Network architecture?

An intuitive question that arises is why we opt for such a full process training, which entails simultaneous training of the $1^{st}$-order-pre-trainer and the $2^{nd}$-order-improver with a large 1000-$1^{st}$ dataset and a small 100-$2^{nd}$, rather than directly fine-tuning the $1^{st}$-order-pre-trainer with the 100-$2^{nd}$ dataset for improved $2^{nd}$ order accuracy in various scenarios.

To explore this further, we directly train the $1^{st}$-order-pre-trainer network to determine its ability to comprehend the relationships between the particles and the grid. Simultaneously, we conducted a sweep analysis of the network with varying layers (i.e., layer numbers = 1,2,3) to investigate the influence of network depth. The outcomes, summarized in Table III, does not suggest any significant improvements even as computational time increased linearly with the number of layer blocks. Thus, directly achieving the $2^{nd}$ order accuracy from the raw particle and grid information seems difficult technically.

Based on the above results, we seek to construct neural networks that indirectly achieve higher order accuracy. The gradual improvements are summarized in the following three steps:

TABLE III
CHARGE DENSITY DEVIATION WITH DIFFERENT LAYER NUMBER AND DATA SIZE

| Layer number | Training Dataset (tested on dataset 10-$2^{nd}$) | |
|---|---|---|
| | 100-$2^{nd}$ | 1000-$2^{nd}$ |
| N=1 | 13.1% | 15.7% |
| | 15.1% | 18.8% |
| | 17.4% | 21.3% |
| (average) | (15.2%) | (16.4%) |
| N=2 | 16.8% | 17.8% |
| | 15.1% | 12.4% |
| | 14.1% | 16.0% |
| (average) | (16.0%) | (15.4%) |
| N=3 | 13.5% | 20.1% |
| | 13.6% | 25.3% |
| | 9.0% | 7.7% |
| (average) | (12.0%) | (17.7%) |

The networks are all trained with $2^{nd}$ charge density distributions after 20000 epochs. $N$ is the number of the encoder-decoder layer. The number in the parentheses is the average of the deviation.

Firstly, we propose training two transformers. The initial one accepts particle coordinates, forecasts the $1^{st}$ order charge density, and conveys it to the second transformer, which in turn predicts the desired $2^{nd}$ order charge density. The comparison, demonstrated as (100-$1^{st}$ + 100-$2^{nd}$) and (1000-$1^{st}$ + 100-$2^{nd}$) in Table IV, reveals some improvements over the results in Table III.

Secondly, we opt to use the discriminator network. The comparison is highlighted as (100-$1^{st}$ + 100-$2^{nd}$ + D) and (1000-$1^{st}$ + 100-$2^{nd}$ + D) in Table IV. As can be observed, there is an optimization of approximately 3~4%.

Thirdly, inspired by the pre-training in GPT [28] , we incorporate an additional pre-training process of the $1^{st}$ order transformer (hence names $1^{st}$-order-pre-trainer). Following this operation, the second transformer (hence names $2^{nd}$-order-improver) can begin its training with results that closely resemble the true $1^{st}$ order charge densities. This comparison is displayed as (1000-$1^{st}$ + 100-$2^{nd}$ + pre-train) and (1000-$1^{st}$ + 100-$2^{nd}$ + D + pre-train) in Table IV. Encouragingly, there is a significant increase of about 5~7% when applying the pre-training process.

Through the above considerations, we finally constructed full-process training in Fig. 4.

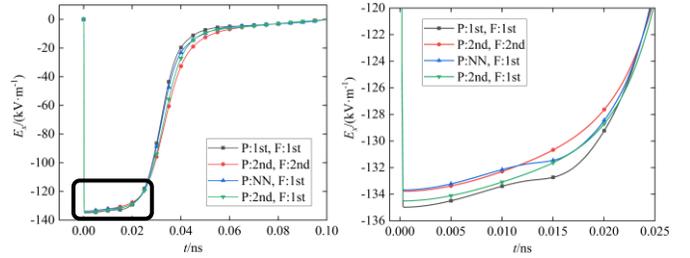

Fig. 10. The x-direction electric field at point (12,12,12) with different particle and field interpolation methods. The black, red, blue, green lines represent the results from JefiPIC with 1st order particle and field interpolation, JefiPIC with 2nd order particle and field interpolation, NN-combined JefiPIC with particle density from NN and 1st order field interpolation from JefiPIC, and JefiPIC with 2nd order particle and 1st order field interpolation.

## TABLE IIIV
### THE CHARGE DENSITY DISTRIBUTION DEVIATE WITH DIFFERENT NETWORK ARCHITECTURE

| Network architecture | Deviation | | | |
|---|---|---|---|---|
| 100-1st + 100-2nd | 7.2% | 12.5% | 12.1% | (10.6%) |
| 100-1st + 100-2nd + D | 8.9% | 7.4% | 6.9% | (7.7%) |
| 1000-1st + 100-2nd | 16.4% | 12.3% | 12.5% | (13.4%) |
| 1000-1st + 100-2nd + D | 6.3% | 17.9% | 14.0% | (9.4%) |
| 1000-1st + 100-2nd + pre-train | 4.6% | 6.3% | 8.5% | (6.5%) |
| 1000-1st + 100-2nd + D + pre-train | 5.5% | 3.8% | 4.5% | (4.6%) |

*100* and *1000* represent the size of the training label, *1st* and *2nd* represent the interpolation order of the charge density, *D* stands for the discriminator network, *pre-train* means the 5000 epochs training of the network 1st-order-pre-trainer. The number in the parentheses is the average of the deviation.

### B. Embedding the trained networks into JefiPIC

Now that we have successfully trained the neural networks on the training sets 1000-1st and 100-2nd, we can incorporate these networks into the PIC code. We can then compare these results with the 2nd order numerical results, allowing us to evaluate the effectiveness of our neural networks.

Fig. 10 displays the x-directional electric fields $E_x$ at point (12,12,12) using different interpolation methods. Here, 'P' represents particle interpolation, 'F' signifies field interpolation, while '1st' and '2nd' denote the interpolation order used in the JefiPIC model. As can be observed, by replacing the conventional particle interpolation with the neural network, the calculated field does not diverge significantly from the numerical results.

Particularly, in the peak value region of the field (encircled by the black rectangle), the electric field generated by the neural network-combined PIC aligns between the results of (P: 2nd, F: 2nd) and (P: 2nd, F: 1st). This indicates that our proposed neural network possesses the capability to stably predict higher order particle density with the attention mechanism.

Table V displays the peak values of the electromagnetic fields at point (12,12,12) using different particle and field interpolation methods, as well as the deviations with respect to the baseline. The baseline is set as the field from JefiPIC with 2nd order field and particle interpolation. According to the data presented, the largest field deviation is observed in the results using 1st order particle and field interpolation.

Interestingly, the field deviation of the NN-combined PIC with 1st order field interpolation is nearly identical to that of the PIC using 2nd order particle and 1st order field interpolation and even smaller at times. This suggests that the incorporation of our neural network can potentially enhance the precision of field prediction.

## TABLE V
### PEAK VALUE OF FIELD BY DIFFERENT INTERPOLATION METHODS.

| Field | 1st Particle 1st Field | 2nd Particle 2nd Field | NN Particle 1st Field | 2nd Particle 1st Field |
|---|---|---|---|---|
| $E_x$/(kV·m⁻¹) | -134.98 (0.90%) | -133.78 | -133.69 (0.07%) | -134.51 (0.55%) |
| $E_y$/(kV·m⁻¹) | -135.37 (1.07%) | -133.94 | -134.59 (0.49%) | -134.54 (0.45%) |
| $E_z$/(kV·m⁻¹) | -49.07 (4.33%) | -51.29 | -49.3 (3.88%) | -50.72 (1.11%) |
| $B_x$/(10⁻⁷ T) | 2.14 (4.46%) | 2.24 | 2.18 (2.68%) | 2.14 (4.46%) |
| $B_y$/(10⁻⁷ T) | -2.07 (8.41%) | -2.26 | -2.15 (4.87%) | -2.16 (4.42%) |

1st and 2nd mean the order of the interpolation. We choose the results from PIC with 2nd order particle and field interpolation as the baseline. The numbers in the parenthesis represent the deviation from the baseline.

This understanding can then be used to provide allocation weights for particle interpolation in PIC simulations. Due to the long-range dependencies in attention, one particle can interact with more grids compared to traditional numerical interpolation, giving our proposed neural network the potential to generate higher order charge density. Furthermore, the softmax activation function maintains charge conservation and helps the network learn the nonlinearity inherent in the interpolation process.

Our neural network was trained using a large number of 1st order data and fine-tuned using a smaller set of 2nd order data and patchGAN. This allows the network to stably predict the charge density distribution with an accuracy approaching 2nd order. It also performed better in terms of training accuracy and time consumption (4.6%, ~2000 s) compared to those trained only with a deeper network architecture and a larger size of 2nd order data (7.7% at best, ~3700 s).

In conclusion, we believe that the application of the attention mechanism in PIC simulations is beneficial. In the future, we aim to study the influence of field interpolation with the help of the attention mechanism.

## V. CONCLUSION

The attention mechanism allows us to understand the similarity between the grid coordinate and particle information.